\documentclass[%
preprint,
 amsmath,amssymb,
 aip
]{revtex4-2}
\usepackage{graphicx}

\begin{document}

\title{Franck-Condon spectra of unbound and imaginary-frequency vibrations via correlation functions: a branch-cut free, numerically stable derivation}

\author{P.~Bryan Changala}
\affiliation{Center for Astrophysics $\vert$ Harvard \& Smithsonian, Cambridge, MA, USA}
\email{bryan.changala@cfa.harvard.edu}
\author{Nadav Genossar}
\author{Joshua H.~Baraban}
\affiliation{Department of Chemistry, Ben-Gurion University of the Negev, Beer Sheva, Israel}

\date{\today}

\begin{abstract}
Molecular electronic spectra can be represented in the time domain as auto-correlation functions of the initial vibrational wavepacket.  We present a derivation of the harmonic vibrational auto-correlation function that is valid for both real and imaginary harmonic frequencies. The derivation rests on Lie algebra techniques that map otherwise complicated exponential operator arithmetic to simpler matrix formulae. The expressions for the zero- and finite-temperature harmonic auto-correlation functions have been carefully structured both to be free of branch-cut discontinuities and to remain numerically stable with finite-precision arithmetic. Simple extensions correct the harmonic Franck-Condon approximation for the lowest-order anharmonic and Herzberg-Teller effects. Quantitative simulations are shown for several examples, including the electronic absorption spectra of F$_2$, HOCl, CH$_2$NH, and NO$_2$.
\end{abstract}

\maketitle

\section{Introduction}
The vibronic structure of optical, photoionization, and photoelectron spectra encode the evolution that nuclear wavepackets undergo following electronic excitation or detachment.
Interpreting the dynamics revealed by such spectra is often aided by first-principles simulations based on two key dynamical simplifications, the Born-Oppenheimer and Franck-Condon (FC) approximations.
Together these imply that the transition dipole between vibrational levels belonging to two different electronic states is proportional to the overlap integral between their respective vibrational wavefunctions. 
If it is sufficiently accurate to further approximate the potential energy surfaces as quadratic expansions about their respective local minima, then the vibrational structure is that of two sets of (mutually displaced, rotated, and scaled) harmonic oscillators, the overlap integrals of which are well known~\cite{Sharp1964,Doktorov1975,Caldwell1980,Fernandez1989}.
As the only information required in this commonly used framework is the equilibrium geometries and quadratic force constants of the initial and final surfaces, there exist a number of popular software packages for computing harmonic FC spectra using this frequency-domain, state-by-state approach~\cite{Dierksen2005a,Rabidoux2016,Li2017,Gozem2021}.
For large molecules, the number of discrete vibrational states that contribute to the spectrum grows rapidly, but intelligent selection algorithms can address this and remain efficient~\cite{Dierksen2005,Santoro2007,Rabidoux2016}.

The frequency-domain approach breaks down when the vertical geometry is highly displaced from a local minimum (if one exists at all) on the final potential energy surface.
A quadratic expansion centered at the minimum would then be a poor approximation at the vertical geometry.
Although an expansion centered at the vertical geometry would be more accurate, the large displacement from the extrapolated effective minimum would still pose a challenge to frequency-domain methods because of the high vibrational state density at the energy of the vertical geometry.
More fundamentally, the Hessian at the vertical geometry is not guaranteed to be positive definite: it may possess negative-curvature (i.e. imaginary-frequency) modes, which generate a continuum of states in the quadratic approximation.
Although one might construct a frequency-domain method that treats such continua directly (or indirectly through non-Hermitian techniques), we adopt a different approach.

The problems of large displacements and negative-curvature modes are naturally addressed by moving to a time-domain picture.
Here the FC spectrum is obtained by the Fourier transform of the auto-correlation function of the initial vibrational wavepacket evolving on the final-state potential energy surface.
Sophisticated time-domain algorithms, such as the multi-configurational time-dependent Hartree (MCTDH) method~\cite{Beck2000a} and Chebyshev propagation~\cite{Chen1999}, can provide highly accurate spectroscopic predictions.
These methods, however, are often computationally expensive, may require special forms of the molecular Hamiltonian and potential energy surfaces, and typically handle unbound modes by imposing outgoing boundary conditions using manually tuned complex absorbing potentials~\cite{Vibok1992,Leforestier1998,Muga2004}.
For spectra that display a broad, continuous FC envelope, only the short-time, and therefore local, dynamics are relevant, motivating a time-domain treatment based on perturbation theory, in the same spirit of standard second-order vibrational perturbation theory (VPT2) for small-amplitude frequency-domain problems~\cite{Luis2004}.
The zeroth-order picture of such an approach would be the harmonic auto-correlation function based on a quadratic expansion of the initial and final potential energy surfaces (both centered about the initial state equilibrium or vertical geometry).
Analytic expressions for the harmonic correlation function are known and indeed have been applied to the general vibrational problem before~\cite{Tannor1982,Ianconescu2004,Baiardi2013,Banerjee2012}.
These formal expressions, however, suffer from two important technical problems.

The first problem is branch-cut discontinuities arising from the periodic motion of bound, real-frequency modes. 
This analytical issue appears in many manifestations of the harmonic oscillator correlation function or propagator (and in the dynamics of oscillators more generally~\cite{Gelabert2000,Horvathy2011}), though it is often overlooked. This may be because there is a simple numerical fix to the problem, which has been adopted in prior applications~\cite{Banerjee2012,Gherib2022,Note1}. Nonetheless, it is of interest to derive an analytical solution to the branch-cut issue for the general, multi-dimensional case, which we present here.
\footnotetext[1]{The procedure is simply to take small enough time-steps to reliably detect a numerical discontinuity in the correlation function.}

The second problem involves unstable finite-precision arithmetic arising from the exponentially diverging motion of unbound, imaginary-frequency modes.
Although the time-domain expressions used in prior applications are formally correct for imaginary-frequency modes by analytic continuation, the details of this procedure require care.
Expressions that are well-behaved numerically for bound, periodic modes may become exponentially divergent for imaginary frequencies. Unitary evolution guarantees that the final correlation function is bounded, but this comes about by cancellation of terms that individually diverge exponentially in time. 
In any practical numerical implementation, guaranteeing these cancellations using finite-precision arithmetic requires a detailed understanding of the analytical structure of the auto-correlation function. Addressing these two issues is a major focus of this paper. 

We first present a derivation of the harmonic correlation function using a Lie-algebraic approach, for both zero and finite temperatures, which enables the clear identification of how both the branch-cut and unstable arithmetic problems arise and how to solve them. Simple extensions beyond the harmonic FC approximation, including the lowest order anharmonic and Herzberg-Teller corrections, are then discussed.
We demonstrate these results with several illustrative examples, highlighting applications to quantitative simulations of the electronic absorption spectra of dissociative surfaces (fluorine, F$_2$, and hypochlorous acid, HOCl), transition state regions (methyleneimine, CH$_2$NH), and highly displaced geometries (nitrogen dioxide, NO$_2$).

\section{Theory}

The harmonic initial state Hamiltonian, with $\hbar = 1$, is 
\begin{align}
H_0 = \sum_i^{n} \frac{1}{2}\omega_i (p_i^2 + q_i^2),
\end{align}
where $q_i$ is the reduced dimensionless normal coordinate of the $i^\mathrm{th}$ vibrational mode ($i = 1,\ldots,n$), $p_i$ is its conjugate momentum, and $\omega_i$ is its harmonic frequency. The dimensionless coordinates and momenta satisfy the canonical commutation relation $[q_j, p_k] = i\delta_{jk}$. In these coordinates, the final state Hamiltonian expanded to second order about the vertical geometry ($q_i = 0$) is
\begin{align}
H &= \sum_i \frac{1}{2} \omega_i p_i^2 + \frac{1}{2} \sum_{ij} K_{ij} q_i q_j + \sum_i G_i q_i \nonumber \\
&= \frac{1}{2} \mathbf{p}^T \mathbf{W} \mathbf{p} + \frac{1}{2} \mathbf{q}^T \mathbf{K} \mathbf{q} + \mathbf{G}^T \mathbf{q} + V_0,\label{eq:Hdef}
\end{align}
where $\mathbf{q}$ and $\mathbf{p}$ are $n \times 1$ column vectors containing the $q_i$ and $p_i$ operators; $\mathbf{W}$ is a diagonal matrix containing $\omega_i$ along the diagonal; and $V_0$, $\mathbf{G}$, and $\mathbf{K}$ are the final state energy, gradient, and Hessian, respectively, evaluated at the vertical geometry. The final state Hamiltonian can be written in terms of its own normal coordinates $\mathbf{Q}$ and momenta $\mathbf{P}$ as
\begin{align*}
    H &= \frac{1}{2}( \mathbf{P}^T \mathbf{\Omega} \mathbf{P} + \mathbf{Q}^T \mathbf{\Sigma}^2 \mathbf{\Omega} \mathbf{Q}) - \frac{1}{2}\mathbf{G}^T \mathbf{K}^{-1} \mathbf{G} + V_0,
\end{align*}
where 
\begin{align*}
    \mathbf{P} &= \mathbf{R}^{-1} \mathbf{p},\\
    \mathbf{Q} &= \mathbf{R}^T (\mathbf{q} - \mathbf{d}),\\
    \mathbf{d} &= -\mathbf{K}^{-1} \mathbf{G},\\
    \mathbf{R} &= \mathbf{W}^{-1/2} \mathbf{L} \mathbf{\Omega}^{1/2}.
\end{align*}
The matrix $\mathbf{L}$ is orthogonal and contains the eigenvectors of the mass-weighted Hessian, i.e.
\begin{align*}
    (\mathbf{W}^{1/2} \mathbf{K} \mathbf{W}^{1/2}) \mathbf{L} = \mathbf{L} \mathbf{Z}
\end{align*}
where $\mathbf{Z}$ contains the real eigenvalues $z_k$ along its diagonal. These may be positive or negative and equal the square of the final-state harmonic frequencies, which we take to be either positive real or negative imaginary. That is, let $z_k^{1/2} \equiv \sigma_k \Omega_k$, where $\Omega_k = \vert z_k\vert^{1/2}$ and $\sigma_k = 1$ if $z_k$ is positive and $\sigma_k = -i$ is $z_k$ is negative. The matrices $\mathbf{\Sigma}$ and $\mathbf{\Omega}$ are diagonal, containing $\sigma_k$ and $\Omega_k$, respectively. The matrices $\mathbf{\Lambda}_\pm$, which relate the canonical creation and annihilation operators of the initial and final states, will appear often below. They are defined as 
\begin{align*}
    \mathbf{\Lambda}_\pm &= \mathbf{W}^{-1/2} \mathbf{L}\mathbf{\Sigma}^{1/2} \mathbf{\Omega}^{1/2} \pm \mathbf{W}^{1/2} \mathbf{L} \mathbf{\Sigma}^{-1/2} \mathbf{\Omega}^{-1/2}.
\end{align*}

\subsection{The zero-temperature correlation function}

For a zero-temperature spectrum, we are interested in calculating the vacuum auto-correlation function,
\begin{align*}
C_0(t) = \langle 0 \vert e^{-i H t} \vert 0 \rangle,
\end{align*}
where $\vert 0 \rangle$ is the ground vibrational state of the initial state Hamiltonian, $H_0$. (The finite-temperature case will be considered in Section~\ref{sec:finiteT}.) The FC spectral density is related to the auto-correlation function by a Fourier transform,
\begin{align}
S_0(\omega) = \frac{1}{2\pi} \int_{-\infty}^{\infty} dt\, e^{i(\omega + E_0) t} C_0(t),\label{eq:specFT}
\end{align}
where $\omega$ equals the transition frequency from the $\vert 0 \rangle$ initial state, which has zero-point energy $E_0 = \sum_k \omega_k/2$.

Our approach to calculating $C_0$ is based on disentangling the exponential propagator. First, we recognize that a harmonic Hamiltonian can always be cast as a general quadratic polynomial in the raising and lowering operators, $a_i^\dagger$ and $a_i$, i.e.~as 
\begin{align}
    -i H t = Q \equiv A_{ij}(a^\dagger_i a_j + \frac{1}{2}\delta_{ij}) + B_{ij} a^\dagger_i a^\dagger_j + C_{ij} a_i a_j + f_i a_i^\dagger + g_i a_i + h,
\end{align}
where repeated indices are implied summations and the $n \times n$ matrices $\mathbf{B}$ and $\mathbf{C}$ are symmetric. (For Hermitian $H$, furthermore, $\mathbf{B} = \mathbf{C}$ and $\mathbf{A}$ is symmetric.) The space of quadratic operators is closed under commutation,
\begin{align*}
    [Q,Q'] = Q'',
\end{align*}
thus forming a Lie algebra. This guarantees that exponential quadratic operators can, in principle, be disentangled via the Baker–Campbell–Hausdorff (BCH) formula as
\begin{align}
    e^Q &= e^{ A_{ij}(a^\dagger_i a_j + \frac{1}{2}\delta_{ij}) + B_{ij} a^\dagger_i a^\dagger_j + C_{ij} a_i a_j + f_i a_i^\dagger + g_i a_i + h} \nonumber \\
    &= e^{h'}e^{f'_i a_i^\dagger} e^{B'_{ij} a_i^\dagger a_j^\dagger} e^{A'_{ij}(a_i^\dagger a_j + \delta_{ij}/2)} e^{C'_{ij} a_i a_j} e^{g'_i a_i}.\label{eq:expQ_disentangled}
\end{align}
The order of the terms in the argument of the exponent in the first line of this equation is arbitrary, but the order of the individual factors in the second line, Eq.~\ref{eq:expQ_disentangled}, is non-trivial because they do not commute.
Assuming the parameters of this particular disentangled form are known, then $C_0$ is simple to evaluate,
\begin{align}
    C_0(t) &=  \langle 0 \vert e^{-i H t} \vert 0 \rangle \nonumber \\
            &= \langle 0 \vert e^{h'}e^{f'_i a_i^\dagger} e^{B'_{ij} a_i^\dagger a_j^\dagger} e^{A'_{ij}(a_i^\dagger a_j + \delta_{ij}/2)} e^{C'_{ij} a_i a_j} e^{g'_i a_i} \vert 0 \rangle \nonumber \\
            &= e^{h'} \langle 0 \vert  e^{A'_{ij}(a_i^\dagger a_j + \delta_{ij}/2)} \vert 0 \rangle \nonumber \\
            &= e^{h'} e^{\text{Tr}[\mathbf{A}'/2]} \nonumber \\
            &= e^{h'} \det[ e^{\mathbf{A'}/2} ]\label{eq:C0val},
\end{align}
where we have made repeated use of the fact that $a_i \vert 0 \rangle = \langle 0 \vert a_i^\dagger = 0$. The problem is now to determine the (time-dependent) disentangled coefficient $h'$ and the determinant of $\exp[\mathbf{A}'/2]$ given the original entangled exponential operator.

To proceed, we use a faithful matrix representation of the quadratic Lie algebra~\cite{Gilmore2008}. Consider the $(2n+2) \times (2n+2)$ matrix, $M$, which is constructed from the quadratic operator coefficients as
\begin{align}
    M &= \left[ \begin{array}{cccc} 0 & \mathbf{g}^T & \mathbf{f}^T & -2h \\ 0 & \mathbf{A} & 2\mathbf{B} & -\mathbf{f} \\ 0 & -2\mathbf{C} & -\mathbf{A}^T & \mathbf{g} \\ 0 & 0 & 0 &0  \end{array}\right].\label{eq:Mdef}
\end{align}
It is straightforward to show that this matrix map of the quadratic operators satisfies the same commutation rules. 
That is, if we let $M$, $M'$, and $M''$ be the matrices mapped to by the operators $Q$, $Q'$, and $Q''$, then $[M,M'] = M''$ if and only if $[Q, Q'] = Q''$.  
The faithful matrix representation is useful because BCH formulas can be derived with them via direct matrix exponentiation and multiplication rather than manipulating the exponential operators themselves~\cite{Gilmore1974}. 
It turns out that the matrices of the form $M$ have a relatively straightforward matrix exponential, so that the disentangling rule, Eq.~\ref{eq:expQ_disentangled}, can be expressed as a $(2n+2)\times(2n+2)$ matrix equation. The full derivation is presented in Appendix~\ref{app:disentangle}.
The two main results needed to evaluate the correlation function are 
\begin{align}
e^{\mathbf{A}'} &= 4 \left( \mathbf{\Lambda}_+ e^+ \mathbf{\Lambda_+}^T - \mathbf{\Lambda}_- e^- \mathbf{\Lambda}_-^T \right)^{-1}\label{eq:expAp}
\end{align}
and
\begin{align}
    h' &= -i t V_0 + \frac{t^2}{16} \mathbf{G}^T (\mathbf{\Lambda}_+ - \mathbf{\Lambda}_-)\mathbf{\Gamma}(\mathbf{\Lambda}_+ - \mathbf{\Lambda}_-)^T \mathbf{G},\label{eq:hp}
\end{align}
where
\begin{align}
\mathbf{\Gamma} &= (\zeta^+ - \zeta^-) -  (\mathbf{\Lambda}_+ \eta^+ + \mathbf{\Lambda}_- \eta^-)^T (\mathbf{\Lambda}_+ e^+ \mathbf{\Lambda}_+^T - \mathbf{\Lambda}_- e^- \mathbf{\Lambda}_-^T)^{-1} (\mathbf{\Lambda}_+ \eta^+ + \mathbf{\Lambda}_- \eta^-),\label{eq:Gamma}
\end{align}
and
\begin{align*}
e^{\pm} &= e^{\pm i t \mathbf{\Sigma} \mathbf{\Omega}}, \\
\eta^\pm &= \frac{ e^{\pm i t \mathbf{\Sigma} \mathbf{\Omega}} - 1}{\pm i t \mathbf{\Sigma}\mathbf{\Omega}},\\
\zeta^\pm &= \frac{ e^{\pm i t \mathbf{\Sigma}\mathbf{\Omega}} \mp i t \mathbf{\Sigma}\mathbf{\Omega} - 1}{-t^2 (\mathbf{\Sigma \Omega})^2}.
\end{align*}

Let us first focus on the matrix $\mathbf{\Gamma}$, which contributes to $h'$. The elements in the diagonal positive-frequency matrices $e^+$, $\eta^+$, and $\zeta^+$ that correspond to modes with imaginary frequencies ($\sigma_k = -i$) are exponentially divergent as $t\rightarrow \infty$. These divergences, however, ultimately cancel in $\mathbf{\Gamma}$ itself, which we can see by assuming the positive-frequency terms dominate the negative-frequency terms at large time,
\begin{align*}
    \mathbf{\Gamma} &\rightarrow \zeta^+ - \eta^+ \mathbf{\Lambda_+}^T (\mathbf{\Lambda}_+ e^+ \mathbf{\Lambda}_+^T)^{-1} \mathbf{\Lambda}_+\eta^+\\
    &=\zeta^+ - \eta^+ e^- \eta^+\\
    &= \zeta^+ - \eta^+ \eta^-\\
    &= \zeta^+ - (\zeta^+ + \zeta^-)\\
    &= -\zeta^-.
\end{align*}
The cancellation of exponentially large intermediate quantities implies that the naive evaluation of $\mathbf{\Gamma}$ using finite-precision floating point arithmetic directly as written in Eq.~\ref{eq:Gamma} may be unstable and inaccurate.
Indeed, numerical tests quickly confirm this issue, suggesting that it is necessary to explicitly remove the divergent intermediates.

To do so, we first inspect various factorizations of the following matrix inverse, which occurs frequently,
\begin{align*}
    \left( \mathbf{\Lambda}_+ e^+ \mathbf{\Lambda_+}^T - \mathbf{\Lambda}_- e^- \mathbf{\Lambda}_-^T \right)^{-1} &= \left(1 - \mathbf{\Lambda}'_+ e^- \mathbf{\Lambda}_+^{-1}\mathbf{\Lambda}_- e^- \mathbf{\Lambda}_-^T \right)^{-1} \mathbf{\Lambda}'_+ e^- \mathbf{\Lambda}_+^{-1} \\
    &= \mathbf{\Lambda}'_+\left(1 -  e^- \mathbf{\Lambda}_+^{-1}\mathbf{\Lambda}_- e^- \mathbf{\Lambda}_-^T \mathbf{\Lambda}'_+ \right)^{-1}  e^- \mathbf{\Lambda}_+^{-1} \\ 
    &= \mathbf{\Lambda}'_+ e^- \left(1 -  \mathbf{\Lambda}_+^{-1}\mathbf{\Lambda}_- e^- \mathbf{\Lambda}_-^T \mathbf{\Lambda}'_+ e^- \right)^{-1} \mathbf{\Lambda}_+^{-1} \\ 
    &= \mathbf{\Lambda}'_+ e^- \mathbf{\Lambda}_+^{-1} \left(1 -  \mathbf{\Lambda}_- e^- \mathbf{\Lambda}_-^T \mathbf{\Lambda}'_+ e^- \mathbf{\Lambda}_+^{-1} \right)^{-1},
\end{align*}
where $'$ denotes the transpose-inverse. In each case, the factorization yields a numerically stable matrix inverse and provides another exponentially damped factor on its right or left. After expanding all terms in Eq.~\ref{eq:Gamma}, one can choose the right- and left-sided forms for each as necessary to cancel the various $\eta^+$ and $\zeta^+$ factors. After some straightforward, if tedious, algebra, the result is 
\begin{align}
    \mathbf{\Gamma} &=  -\eta^- \mathbf{\Lambda}_+^{-1} \mathbf{\Lambda}_- e^- \mathbf{\Lambda}_-^T \mathbf{\Lambda}'_+ (1 -   e^- \mathbf{\Lambda}_+^{-1} \mathbf{\Lambda}_- e^- \mathbf{\Lambda}_-^T \mathbf{\Lambda}'_+)^{-1} \eta^-  \nonumber \\ 
    &\qquad -\left[\eta^- \mathbf{\Lambda}_-^T (1 -  \mathbf{\Lambda}'_+ e^- \mathbf{\Lambda}_+^{-1} \mathbf{\Lambda}_- e^- \mathbf{\Lambda}_-^T)^{-1} \mathbf{\Lambda}'_+  \eta^- + \text{transpose}\right] \nonumber \\
    &\qquad - 2\zeta^- - \eta^-  \mathbf{\Lambda}_-^T(1 -  \mathbf{\Lambda}'_+ e^- \mathbf{\Lambda}_+^{-1} \mathbf{\Lambda}_- e^- \mathbf{\Lambda}_-^T)^{-1} \mathbf{\Lambda}'_+ e^- \mathbf{\Lambda}_+^{-1} \mathbf{\Lambda}_- \eta^-. \label{eq:Gammanorm}
\end{align}
The non-commuting nature of matrix multiplication makes this expression appear rather cumbersome. Nonetheless, all exponentially divergent intermediates have been removed, and it can be numerically evaluated accurately.

We now proceed to the second factor of $C_0(t)$, the determinant of $\exp[\mathbf{A}'/2]$. The disentangling formula yields only the matrix exponential of $\mathbf{A}'$ itself (Eq.~\ref{eq:expAp}). The determinants of these two matrices are, of course, related by a square root, but the relation $\det[ e^{\mathbf{A'}/2} ] = \left(\det[ e^{\mathbf{A'}} ]\right)^{1/2}$ requires careful consideration of the branch-cut choice implicit in the square root in the right-hand side. If $(\cdots)^{1/2}$ is taken to be the principal square root, which we denote with $\sqrt{\cdots}$, then each time $\det[ e^{\mathbf{A'}}]$ crosses the negative real axis in the complex plane, its square root would experience a sign-reversing branch-cut discontinuity. 

To examine the problem more closely, we first consider a single vibrational mode, in which case the matrix expressions reduce back to simple scalar ones. Eq.~\ref{eq:expAp} simplifies to
\begin{align*}
    e^{A'} &= \left[ \cos(\sigma \Omega t) + i a \sin(\sigma \Omega t) \right]^{-1},
\end{align*}
where $a = (\omega^2 + \sigma^2 \Omega^2)/2\omega \sigma \Omega$. For a positive-curvature, real-frequency mode, $\sigma = 1$ and $a \geq 1$, so the expression in the brackets orbits an elliptical trajectory in the complex plane with frequency $\Omega$.
At values of $\Omega t = \pi, 3\pi, 5\pi, \ldots$, the trajectory crosses the negative real axis and the principal square-root $\sqrt{e^{A'}}$ changes sign discontinuously.
The problem, of course, is related to the leading $e^{i \Omega t}$ like behavior, for which it is clear that $e^{i \Omega t/2} \neq \sqrt{e^{i \Omega t}}$ for some $t$.
This suggests that the solution is to factor out the global $e^{i \Omega t}$ dependence of the complex orbit as 
\begin{align}
    e^{A'} = \frac{2}{1+a}e^{-i \sigma \Omega t} \left(1 + \frac{1-a}{1+a} e^{-2i \sigma \Omega t}\right) ^{-1}.\label{eq:expA_fact_1d}
\end{align}
We now take the square-root of each factor separately,
\begin{align}
    e^{A'/2} = \sqrt{\frac{2}{1+a}} e^{ - i \sigma \Omega t / 2} \sqrt{\left(1 + \frac{1-a}{1+a} e^{-2i \sigma \Omega t}\right)^{-1}},\label{eq:cutfree_1d}
\end{align}
where all $\sqrt{\cdots}$ are principal square roots. 
Because $a\geq 1$ (for real-frequency modes), the quantity $(1-a)/(1+a)$ has an absolute value less than unity and the argument of the right-most square root never crosses the negative real axis.
(The quantity in the parentheses makes a simple circular orbit centered about 1.)
Its principal square root thus exhibits no branch-cut discontinuity.
The global phase of the orbit is carried by the pure exponential factor $e^{-i \sigma \Omega t /2}$, which can be calculated directly because $\sigma$ and $\Omega$ are known.
The branch-cut free expression, Eq.~\ref{eq:cutfree_1d}, also holds for a negative-curvature, imaginary-frequency mode, which has $\sigma = -i$.
In this case, $a$ is a purely imaginary number, so that $(1-a)/(1+a)$ is a complex number of unit modulus, and the argument of the square root again does not cross the negative real axis. (This is assuming $ t \geq 0$. For $t < 0$, one need only calculate $C_0(|t|)$ and recognize that $C_0(-t) = C_0(t)^*$.)

This simple factorization trick must now be generalized to the multidimensional case. Starting from Eq.~\ref{eq:expAp}, we factor the matrix inverse as
\begin{align*}
    e^{\mathbf{A}'} &= 4 ( 1 - \mathbf{\Lambda}'_+ e^- \mathbf{\Lambda}_+^{-1} \mathbf{\Lambda}_- e^- \mathbf{\Lambda}_-^T)^{-1} \mathbf{\Lambda}'_+ e^- \mathbf{\Lambda}_+^{-1}
\end{align*}
and then calculate its determinant. Recognizing that $\det(\mathbf{A}\mathbf{B}) = \det(\mathbf{A}) \det(\mathbf{B})$ and $\det(\mathbf{A}) = \det(\mathbf{A}^T)$, we have
\begin{align*}
    \det (e^{\mathbf{A}'}) &= \det(\mathbf{\Lambda}_+/2)^{-2}  e^{-it \text{Tr}[\mathbf{\Sigma \Omega}]} \det( 1 - \mathbf{\Lambda}'_+ e^- \mathbf{\Lambda}_+^{-1} \mathbf{\Lambda}_- e^- \mathbf{\Lambda}_-^T)^{-1} .
\end{align*}
These three factors correspond one-to-one with those of Eq.~\ref{eq:expA_fact_1d}, and we need a branch-cut-free square root of each. As in the one-dimensional case, the first factor, $\det(\mathbf{\Lambda}_+/2)^{-2}$, is simply a constant and poses no problems. The second factor, $e^{-it \text{Tr}[\mathbf{\Sigma \Omega}]} = \prod_k e^{-i t \sigma_k \Omega_k}$, is the product of the global phase factor for each normal mode. For real-frequency modes, this will be a periodic, complex phase. For imaginary-frequency modes, this will be an exponential damping term. As all $\sigma_k$ and $\Omega_k$ are known, $e^{-it \text{Tr}[\mathbf{\Sigma \Omega}]/2}$ can be calculated directly. The third factor requires more care. 
At this point in the one-dimensional case, we could directly proceed to the principal square root. However, in the multidimensional case, the determinant in the third factor can still cross the negative real axis, and in general does. A further factorization is necessary, and given that this is a determinant, it is natural to consider the spectrum of the matrix argument. Let $\lambda_k$ ($k = 1, \ldots, n$) be the eigenvalues of $(1 - \mathbf{\Lambda}'_+ e^- \mathbf{\Lambda}_+^{-1} \mathbf{\Lambda}_- e^- \mathbf{\Lambda}_-^T)$. Its determinant is then $\lambda_1 \lambda_2 \cdots \lambda_n$. We assert that each eigenvalue individually does not cross the negative real axis. Therefore, a branch-cut-free square root of the determinant can be constructed by multiplying the principal square root of each eigenvalue respectively. The final determinant is then
\begin{align}
    \det (e^{\mathbf{A}'/2}) = \det(\mathbf{\Lambda}_+/2)^{-1}  e^{-it \text{Tr}[\mathbf{\Sigma \Omega}]/2} \prod_k \sqrt{\lambda_k^{-1}}.\label{eq:cutfree_nd}
\end{align}
In summary, the numerically stable, discontinuity-free vacuum auto-correlation function $C_0(t)$ is given by Eqs.~\ref{eq:C0val}, \ref{eq:hp}, \ref{eq:Gammanorm}, and \ref{eq:cutfree_nd}.

\subsection{The finite-temperature correlation function\label{sec:finiteT}}
The general finite-temperature correlation function replaces the vacuum expectation value with a thermal-average trace, 
\begin{align}
C(t, \beta) &= \frac{1}{Z_0(\beta)} \text{Tr}\left[ e^{(+it-\beta)H_0} e^{-it H} \right],\label{eq:Cthermal}
\end{align}
where $\beta = 1/kT$ and $Z_0(\beta) = \text{Tr}[e^{-\beta H_0}]$, the ground state partition function.
The finite-temperature spectrum is related to the correlation function by the same Fourier transform as for the zero-temperature case, Eq.~\ref{eq:specFT}, with $C_0(t)$ replaced by $C(t,\beta)$. In the zero-temperature case, we found a form of the exponential operator in which the vacuum expectation value was simple to evaluate (the disentangled form) and then used the Lie algebra matrix representation to express the exponential in that form. The finite-temperature case proceeds in a parallel fashion. We first find a form in which the trace is simple to evaluate (the diagonalized form, considered below), and then use the same matrix representation to express the exponential product in Eq.~\ref{eq:Cthermal} in that form.

Consider first the trace of a purely quadratic exponential operator, 
\begin{align*}
   \text{Tr}\big[ \exp[A_{ij}(a^\dagger_i a_j + \delta_{ij}/2) + B_{ij}(a_i a_j + a^\dagger_i a^\dagger_j)]\big],
\end{align*}
where we have assumed that $B_{ij} = C_{ij}$ and $A_{ij}$ is symmetric.
The trace is invariant to similarity transformations. Thus, a series of rotation and scaling transformations can diagonalize this operator to
\begin{align*}
    \text{Tr}[e^{\omega'_i(a^\dagger_i a_i + 1/2)}] &= \prod_i \frac{e^{\omega'_i / 2}}{1-e^{\omega'_i}},
\end{align*}
where $\omega_i'^2$ are the eigenvalues of $\mathbf{G} \mathbf{F}$, for $\mathbf{G} = \mathbf{A} - 2 \mathbf{B}$ and $\mathbf{F} = \mathbf{A} + 2 \mathbf{B}$. The parameters of the similarity transformations themselves are not needed. The square of this trace is 
\begin{align*}
    \text{Tr}[\cdots]^2 &= \prod_i \frac{1}{2} \frac{1}{\cosh \omega_i' - 1},
\end{align*}
which is invariant to the branch choice of $\omega'_i$. If we let $\mu$ denote the $2n \times 2n$ matrix,
\begin{align*}
    \mu &= \left[ \begin{array}{cc} \mathbf{A} & 2\mathbf{B} \\ -2\mathbf{B} & -\mathbf{A} \end{array}\right],
\end{align*}
then it can be shown that the squared trace can also be expressed as
\begin{align*}
    \text{Tr}[\cdots]^2 &= (-1)^n \det\left[ e^\mu - \mathbf{1} \right]^{-1}.
\end{align*}
Including gradient terms and a constant energy offset adds a contribution of the form $g_i (a_i + a_i^\dagger) + h$ to the exponential argument. The gradient terms can be eliminated using displacement similarity transformations (which again leave the trace invariant), modifying the total constant term to
\begin{align}
    h' = -\mathbf{g}^T \left( \mathbf{A} + 2 \mathbf{B} \right) ^{-1} \mathbf{g} + h.\label{eq:hp_finiteT}
\end{align}
Thus, the squared trace of a general symmetric quadratic operator is
\begin{align*}
    \text{Tr}[\cdots]^2 &= (-1)^n \det \vert e^\mu - \mathbf{1} \vert ^{-1} \times e^{2h'}.
\end{align*}

The exponential product in Eq.~\ref{eq:Cthermal} must now be recast into a single exponential operator, which can then be ``diagonalized'' in the sense above. The trace is first symmetrized as
\begin{align}
\text{Tr}\left[ e^{(+it-\beta)H_0} e^{-it H_1} \right] &= \text{Tr}\left[ e^{-\tau H_0} e^{-it H_1} \right] \nonumber \\
&= \text{Tr}\left[ e^{-\tau H_0/2 } e^{-it H_1} e^{-\tau H_0 /2 }\right],\label{eq:splittrace}
\end{align}
where $\tau = -it + \beta$. From here, the exponential product is carried out in the finite matrix representation in a similar fashion as the zero-temperature case. The full derivation is shown in Appendix~\ref{app:finiteT}, which includes the somewhat lengthy expressions for the final branch-cut free and numerically stable result for $C(t, \beta)$. The result is also well-behaved in the $T\rightarrow 0$ ($\beta \rightarrow \infty$) limit.  
\subsection{Simple anharmonic corrections}\label{sec:cubic}

The anharmonicities of the initial and final state potential energy surfaces both affect the auto-correlation function, the former by modifying the initial wavefunctions and partition function and the latter by modifying the propagation. Considering only the zero-temperature limit, these can be expanded in a perturbation series,
\begin{align*}
    \vert \Psi \rangle &= \vert 0 \rangle + \lambda \vert \Psi^{(1)} \rangle + \lambda^2 \vert \Psi^{(2)} \rangle \cdots \\
    e^{-i (H+V') t} &= e^{-i H t}(1 + \lambda u^{(1)}(t) + \lambda^2 u^{(2)}(t) \cdots),
\end{align*}
where $\lambda$ is a dummy order-sorting parameter and $V'$ is the anharmonic contribution to the upper state surface. The anharmonic term of the initial state surface, $V$, gives corrections to the ground state $\vert \Psi \rangle$.

Formally, the complete first-order correction to the correlation function is
\begin{align*}
    \lambda C_0^{(1)}(t) &= \lambda\left( \langle 0 \vert e^{-iHt} \vert \Psi^{(1)} \rangle + \langle \Psi^{(1)} \vert e^{-iHt} \vert 0 \rangle + \langle 0 \vert e^{-i H t} u^{(1)}(t) \vert 0 \rangle \right) \\
    &= \lambda \langle 0 \vert e^{-i H t} \left( 2 \vert \Psi^{(1)} \rangle + u^{(1)}(t)\vert 0 \rangle \right).
\end{align*}
The first-order initial state term $\vert \Psi^{(1)} \rangle$ can be computed straightforwardly using time-independent vibrational perturbation theory, expanded in terms of the zeroth-order harmonic oscillator states,
\begin{align*}
    \vert \Psi^{(1)} \rangle = \sum_\mathbf{m} c_\mathbf{m} \vert \mathbf{m} \rangle,
\end{align*}
where $\mathbf{m} = \left\{m_1, m_2, \ldots, m_n\right\} $ are the $n$ quantum numbers of the zeroth-order states. The harmonic cross-correlation functions
\begin{align*}
    \langle \mathbf{m} \vert e^{-i H t} \vert 0 \rangle,
\end{align*}
can in turn be computed in terms of $C_0(t)$ using the multidimensional recurrence relations derived by Fern{\'{a}}ndez and Tipping~\cite{Fernandez1989,Note2}. On the other hand, the final-state propagator correction $u^{(1)}(t)$ requires a significantly more complex time-dependent treatment, including the renormalization of secular terms~\cite{Iso2018}, which is beyond the scope of this paper. As will be seen below, however, the time-independent initial state term is in many cases the dominant first-order contribution, and the time-dependent final state term can be neglected.\footnotetext[2]{These recurrence relations exhibit similar numerical instabilities as $C_0(t)$ for imaginary-frequency modes, and the same factorizations as those used above provide for stable finite-precision evaluation. We note that the recurrence relations, which provide the ratio of $\langle \mathbf{m} \vert e^{-i H t} \vert \mathbf{n} \rangle$ with respect to $C_0(t)$, can be used to derive a simple differential equation for $C_0(t)$ itself. This differential equation has the feature of having no branch-cut discontinuities. To solve for $C_0(t)$ via numerical integration, however, a sufficiently small integration time-step must be used, leading to the same practical circumstance as taking small time-steps to manually correct sign changes in the direct approach.}

We anticipate this scenario to occur when there is a large upper state gradient at the vertical geometry. In this case, the leading anharmonic corrections are due to slight changes in the average position $\langle \mathbf{q} \rangle$ of the initial wavefunction $\vert \Psi \rangle$, which are amplified by the upper state gradient and shift the effective vertical energy. Accounting for cubic anharmoncity, the first-order displacement in each normal coordinate can be evaluated using standard harmonic oscillator matrix elements as
\begin{align*}
    \langle q_i \rangle &= -\frac{1}{4 \omega_i} \sum_j \phi_{ijj},
\end{align*}
where $V = \frac{1}{6} \sum_{ijk} \phi_{ijk} q_i q_j q_k$ are the ground-state anharmonic terms. The approximate spectral shift is
\begin{align}
    \Delta E_\text{cubic} &\approx \mathbf{G} \cdot \langle \mathbf{q} \rangle \nonumber \\
    &= -\frac{1}{4} \sum_{ij} \frac{G_i \phi_{ijj}}{\omega_i}, \label{eq:dEapprox}
\end{align}
where $\mathbf{G}$ is the upper-state gradient at the vertical geometry.

\subsection{Linear Herzberg-Teller corrections}

Accounting for the vibrational dependence of the electronic transition dipole moment requires consideration of the full dipole auto-correlation function, which for the zero-temperature case is
\begin{align}
    D_0(t) &= \langle 0 \vert \mu(\mathbf{q}) e^{-i H t} \mu(\mathbf{q}) \vert 0 \rangle,\label{eq:dipoleauto}
\end{align}
where we assume the transition dipole function $\mu(\mathbf{q})$ is real.
The Franck-Condon approximation truncates the dipole function to its value at the initial state equilibrium geometry $\mu_e = \mu(\mathbf{q} = 0)$, so that $D_0(t) \approx \mu_e^2 C_0(t)$. For electronic transitions where $\mu_e = 0$ by symmetry or $\mu(\mathbf{q})$ varies strongly near $\mathbf{q} = 0$, the linear and possibly higher-order $\mathbf{q}$-dependence of $\mu$ (the so-called Herzberg-Teller effect) needs to be accounted for.

Here, we consider just linear dependence,
\begin{align*}
    \mu(\mathbf{q}) = \mu_e + \sum_i \partial_i\mu \, q_i,
\end{align*}
where $\partial_i \mu = \partial \mu / \partial q_i \vert_e$ are the equilibrium transition dipole derivatives. The wavefunction $\mu(\mathbf{q}) \vert 0 \rangle$ can be expanded in terms of the vacuum and one-quantum excited states. The cross-correlation functions of each of these terms are evaluated using the same recurrence relations as in Section~\ref{sec:cubic}. A similar procedure can be used to efficiently combine linear Herzberg-Teller effects with the anharmonic ground state that includes first-order cubic corrections $\vert \Psi^{(1)} \rangle$.

As with the cubic anharmonicity corrections, the linear dipole dependence will be particularly important when there is a large upper state gradient at the vertical geometry. We can estimate the shift to the effective vertical energy by evaluating the average displacement of $\mu\vert 0 \rangle$, which to first order is
\begin{align*}
    \langle q_i \rangle &= \frac{ \langle 0 \vert \mu q_i \mu \vert 0 \rangle}{\langle 0 \vert \mu^2 \vert 0 \rangle}\\
    &\approx \frac{\partial_i \mu}{\mu_e},
\end{align*}
i.e., the fractional dipole derivative. The expected spectral energy shift analogous to Eq.~\ref{eq:dEapprox} is
\begin{align}
    \Delta E_\text{HT} &\approx \mathbf{G} \cdot \langle \mathbf{q} \rangle \nonumber\\
    &= \sum_i G_i \frac{\partial_i \mu}{\mu_e}.\label{eq:dEHT}
\end{align}

\section{Examples}

We now explore various applications of the harmonic correlation function and its low-order anharmonic and Herzberg-Teller corrections as implemented in the \textsc{Nitrogen} Python package~\cite{NITROGEN}. Focus is placed on systems for which the time-domain approach is most appropriate. The correctness of the implementation for bound harmonic calculations has been verified by comparison with frequency-domain programs~\cite{Rabidoux2016} and for unbound problems by comparison with numerical wavepacket propagation on low-dimensional model potentials.

\subsection{Dissociative excited states of F$_2$ and HOCl}

\begin{figure}
\centering
\includegraphics[width=6.0in]{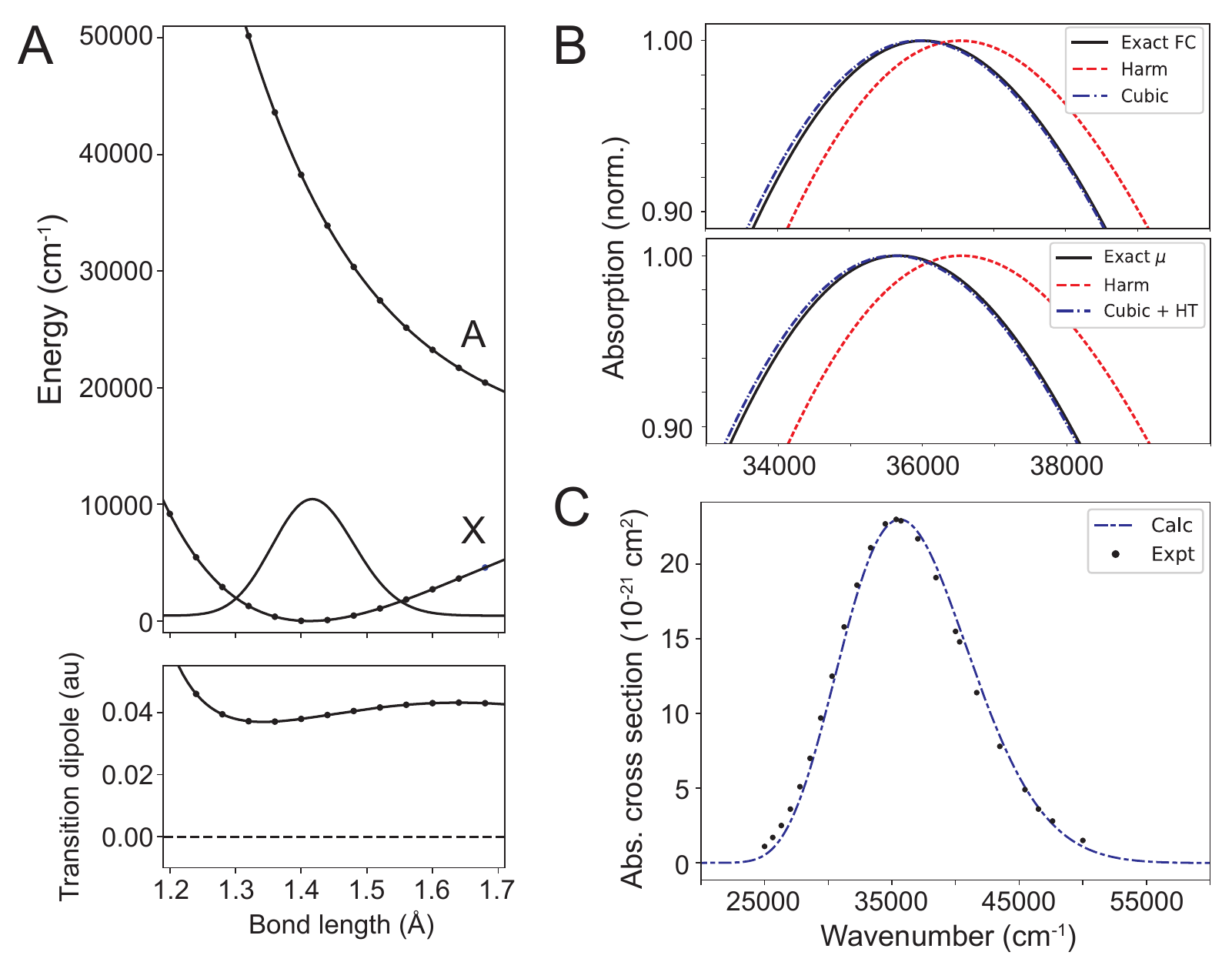}
\caption{The $A$--$X$ band of F$_2$. (A) The potential energy curves (top panel) and transition dipole (bottom panel) of the $X$ and $A$ states show the \textit{ab initio} points (dots) and the fitted curves (solid line). The ground vibrational wavefunction is plotted over the $X$ state curve. (B) The top panel compares the exact Franck-Condon simulation (black, solid) with the harmonic (red, dashed) and first-order cubic (blue, dot-dashed) simulations. The bottom panel shows the exact $\mu(r)$-dependent simulation (black, solid) with the harmonic (red, dashed) and first-order cubic plus linear Herzberg-Teller (blue, dot-dashed) simulations. (C) The calculated first-order cubic plus linear Herzberg-Teller absolute absorption cross sections (solid) are compared with the 298~K experimental spectrum~\cite{Holland1987} (dots).} 
\label{fig:F2curves}
\end{figure}

The absorption spectrum of the dissociative $A$--$X$ transition of F$_2$ provides a simple one-dimensional example to compare the harmonic correlation and its low-order corrections to the corresponding numerically exact results on a given potential energy curve. We calculated the potential energy and transition dipole curves on a grid of 16 points spaced evenly over $r \in [1.2,1.8]$~{\AA} at the frozen-core CCSDT level of theory~\cite{Noga1987} with an aug-cc-pVQZ basis set~\cite{Kendall1992} for the $X$ state and its equation-of-motion (EOM) variant~\cite{Kucharski2001} for the excited $A$ state. (All electronic structure calculations were performed with the \textsc{Cfour} package~\cite{Matthews2020}.) The transition dipole curve was calculated from the geometric mean of the left and right EOM expectation values. Each curve was fitted to an 8$^\text{th}$-order polynomial in the scaled coordinate $x = e^{-r/a}$, with $a = 1.0$~{\AA}, from which derivatives were calculated. The potential energy and transition dipole curves are shown in Fig.~\ref{fig:F2curves}A.

The effects of anharmonicity in the $X$ state surface are illustrated in the top panel of Fig.~\ref{fig:F2curves}B, where the numerically exact FC spectrum is compared against the harmonic simulation and that which includes first-order ground-state cubic corrections. The first-order spectrum agrees closely with the exact result. The shift of the spectral peak is about $-580$~cm$^{-1}$, while the simple cubic-gradient approximation (Eq.~\ref{eq:dEapprox}) predicts $-780$~cm$^{-1}$. The corresponding shifts from the linear Herzberg-Teller corrections are shown in the bottom panel of Fig.~\ref{fig:F2curves}B, where the exact $\mu(r)$-dependent simulation is compared with the harmonic FC simulation and that which includes both first-order cubic and linear Herzberg-Teller corrections. The shift between the simulations with and without Herzberg-Teller corrections is $-360$~cm$^{-1}$. The corresponding estimate of this shift (Eq.~\ref{eq:dEHT}) is $-350$~cm$^{-1}$. 

The experimentally measured 298~K absorption cross sections~\cite{Holland1987} are compared to the first-order cubic/linear Herzberg-Teller simulation in Fig.~\ref{fig:F2curves}C. The absolute absorption cross section is related to the dipole spectral density $D(\omega)$ as
\begin{align*}
    \sigma(\omega) = A \times g_e \times \omega D(\omega),
\end{align*}
where $g_e$ is the electronic degeneracy and $A = 2\pi^2/3\epsilon_0 h c \approx 2.6891 \times 10^{-18}$~cm$^2/(e\,a_0)^{2}$. The $A$ state of F$_2$ has $\Pi$ symmetry, so that $g_e = 2$. The agreement between the calculated spectrum, which has not been scaled or shifted, and the experimental data is excellent.

\begin{figure}[h!]
    \centering
    \includegraphics{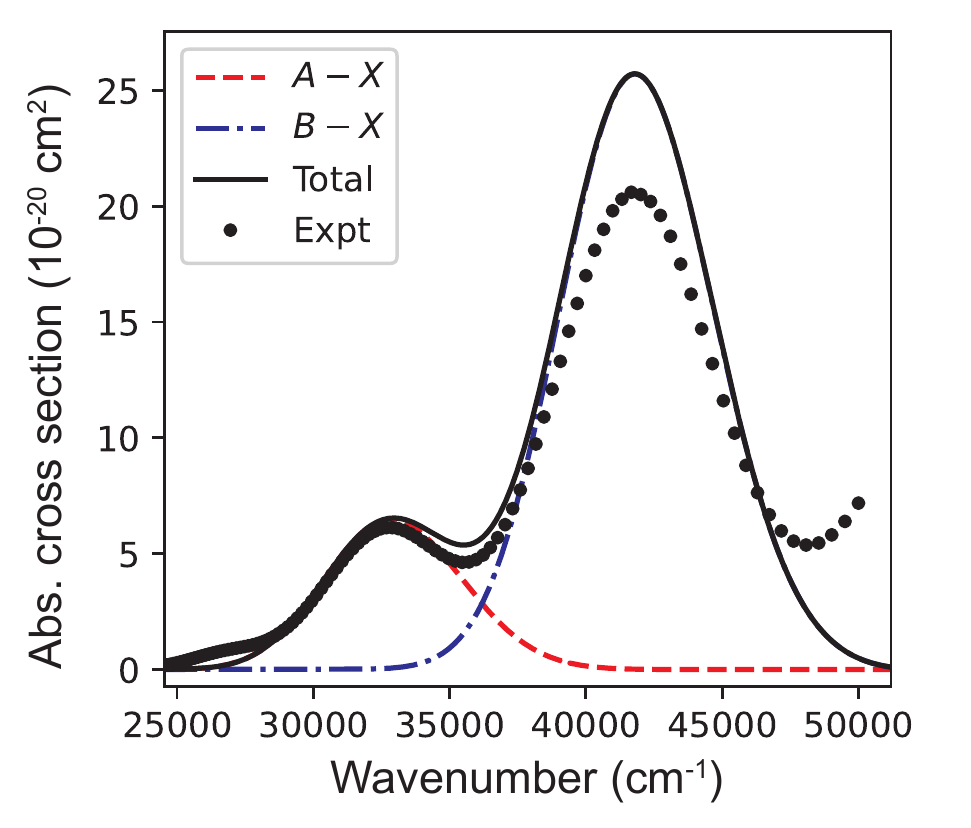}
    \caption{The absorption spectrum of the $\tilde{A}$--$\tilde{X}$ and $\tilde{B}$--$\tilde{X}$ bands of HOCl. The EOM-CCSDT/aug-cc-pVQZ simulation includes first-order cubic and linear Herzberg-Teller corrections. The contributions from the $\tilde{A}$ (red, dashed) and $\tilde{B}$ (blue, dot-dashed) states are shown together with their sum (black, solid). The experimental data points (dots) are at 298~K~\cite{Burkholder2019}.}
    \label{fig:HOClexpt}
\end{figure}

As an example of a dissociative system with bound spectator modes, we examined the absorption spectrum of the $\tilde{A}$ and $\tilde{B}$ states of HOCl, both of which dissociate to OH + Cl. The force constants and transition dipole derivatives were determined by a polynomial fit to a grid of 120 \textit{ab initio} points near the $\tilde{X}$ state equilibrium geometry calculated at the same frozen-core (EOM-)CCSDT/aug-cc-pVQZ level of theory employed for F$_2$. The experimental absorption cross sections~\cite{Burkholder2019,Burkholder1993,Barnes1998} are compared to the simulation with first-order cubic and linear Herzberg-Teller corrections in Fig.~\ref{fig:HOClexpt}.

The single largest discrepancy is the total intensity of the $\tilde{B}$--$\tilde{X}$ band. That is, the error arises primarily from the \textit{ab initio} transition dipole. The accurate shape and breadth of both band systems indicate that the approximate correlation functions and force-fields are accurate. The low-energy shoulder in the experimental spectrum is due to weak absorption from the lowest-lying triplet state~\cite{Barnes1998} and a full treatment of the absolute intensities of this system requires that spin-orbit interactions with the triplet manifold be taken into account~\cite{Minaev1998}, well outside the scope of the present study. We also note that finite-temperature effects, calculated at the harmonic approximation, are minor for these systems at 298~K.

\subsection{Accessing isomerization barriers: CH$_2$NH}

Methyleneimine, CH$_2$NH, is the simplest imine. It has a planar ground-state equilibrium structure~\cite{Pearson1977}, and as in the isoelectronic case of ethylene, relaxes to a non-planar geometry with a dihedral angle of 90$^\circ$ between the CNH and CH$_2$ planes in its first singlet valence excited state~\cite{Bonacic-Koutecky1985,Hazra2004}. The two equivalent non-planar $C_s$ minima of the excited state are connected by a planar $C_{2v}$ transition state with a linear CNH bond angle~\cite{Bonacic-Koutecky1985}. Thus, at the ground-state equilibrium geometry, the excited state is best described as a displaced transition state, possessing a large gradient along the CNH bond angle (initially relaxing towards 180$^\circ$) and imaginary frequencies for both the NH torsion and CH$_2$ out-of-plane angles. Upon $\tilde{A}$--$\tilde{X}$ excitation, the ground state vibrational wavepacket will quickly relax along these degrees of freedom leading to a broad absorption envelope~\cite{Teslja2004}. 

\begin{figure}
    \centering
    \includegraphics{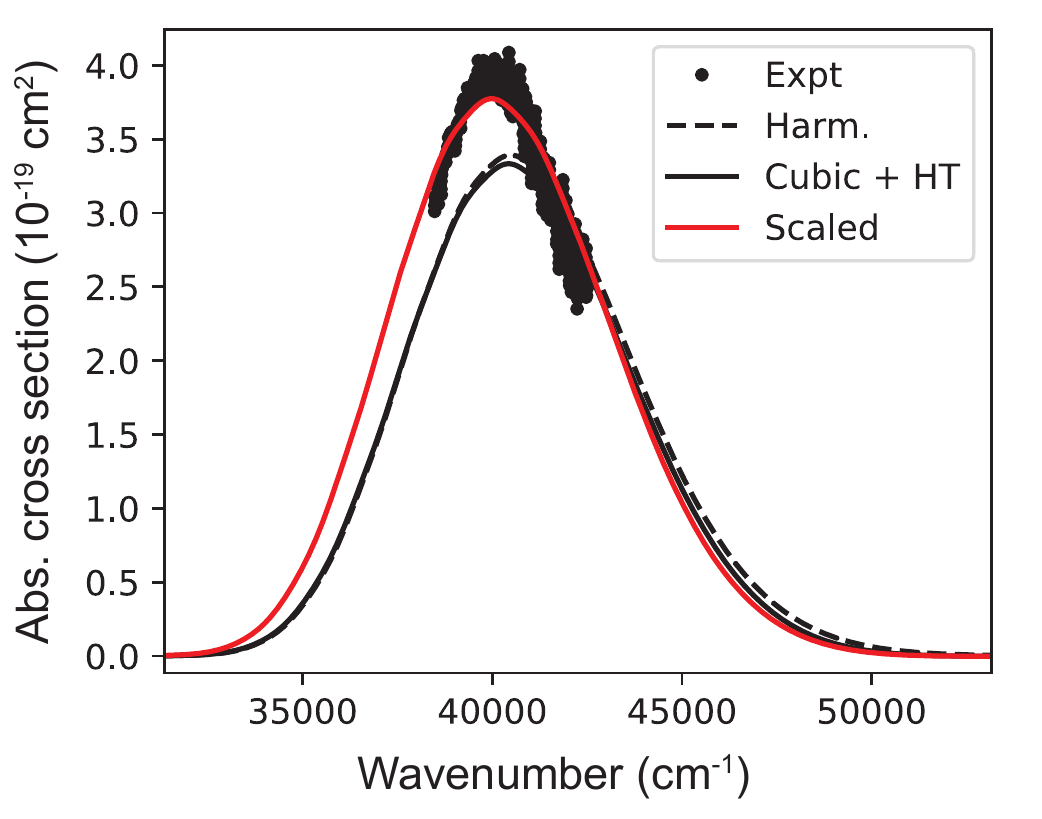}
    \caption{The absorption spectrum of the $\tilde{A}$--$\tilde{X}$ band of CH$_2$NH. The room-temperature experimental spectrum~\cite{Teslja2004} is shown with the harmonic (black, dashed) and first-order cubic/linear Herzberg-Teller-corrected (black, solid) simulations at the EOM-CCSD(T)(a)*/aug-cc-pVQZ level of theory. (Simulated resolution = 1050~cm$^{-1}$~FWHM.)
    A scaled simulation with the vertical energy decreased by 450~cm$^{-1}$ and the 
    transition dipole increased by 7\% is also shown (red, solid). }
    \label{fig:CH2NHexpt}
\end{figure}

To simulate the absorption spectrum, we computed the force fields and transition dipole of the $\tilde{X}$ and $\tilde{A}$ states at the frozen-core EOM-CCSD(T)(a)*/aug-cc-pVQZ level of theory~\cite{Matthews2016}. The harmonic and first-order cubic/linear Herzberg-Teller-corrected simulations are compared to the room-temperature experimental spectrum~\cite{Teslja2004} in Fig.~\ref{fig:CH2NHexpt}. The simulation is slightly too high in energy and too low in intensity. Given the relatively small effect of the cubic and Herzberg-Teller corrections, we speculate that this error arises primarily from the vertical electronic energy and equilibrium transition dipole. Indeed, modest shifts of $-450$~cm$^{-1}$ to the vertical energy and $+7\%$ to the transition dipole bring the simulation in close agreement with the measurements. 

\subsection{Highly displaced vertical geometry: NO$_2$}

\begin{figure}
    \centering
    \includegraphics[width=6in]{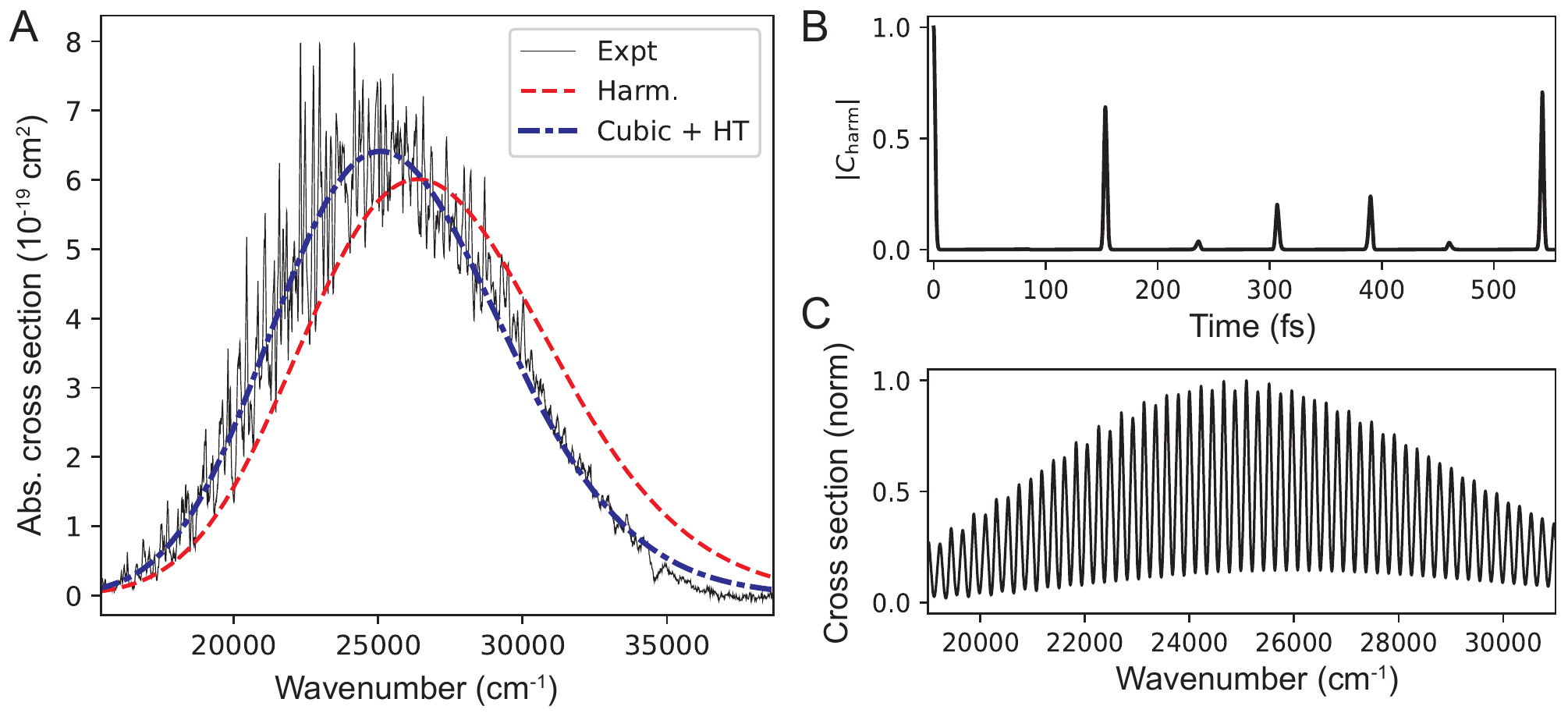}
    \caption{The absorption spectrum of the $\tilde{A}$--$\tilde{X}$ band of NO$_2$. (A) The experimental spectrum~\cite{Bogumil2003,Keller-Rudek2013} at 203~K (thin solid) is shown with the harmonic (red, dashed) and first-order cubic/linear Herzberg-Teller-corrected (blue, dot-dashed) simulations (FWHM = 500~cm$^{-1}$). The absolute value of the harmonic auto-correlation function shows periodic wavepacket revivals (B), leading to regular vibrational progressions in a high-resolution (FWHM = 85~cm$^{-1}$) simulation (C). }
    \label{fig:NO2expt}
\end{figure}

NO$_2$ exhibits a large change in geometry upon transition to its first excited electronic state, with an equilibrium $\angle$ONO bond angle of 134$^\circ$ in the $\tilde{X}$ state and 102$^\circ$ in the $\tilde{A}$ state~\cite{Ndengue2021}, leading to an extended Franck-Condon envelope. 
We simulated the $\tilde{A}-\tilde{X}$ absorption spectrum using EOMIP-CCSD(T)(a)*/aug-cc-pVQZ force fields based on an NO$_2$ anion reference wavefunction. We assume a vertical transition dipole of $0.302$~a.u.~based on the high-level MRCI calculations of Ndengu{\'{e}} \textit{et al.}~\cite{Ndengue2021} and supplement this with fractional dipole derivatives calculated at the EOMEE-CCSD/aug-cc-pVQZ level using the neutral $\tilde{X}$ state reference wavefunction. Figure~\ref{fig:NO2expt}A compares the experimental spectrum obtained at 203~K~\cite{Bogumil2003,Keller-Rudek2013} with the harmonic and ground-state-cubic/linear Herzberg-Teller spectra, simulated with a Gaussian FWHM resolution of 500~cm$^{-1}$. As with previous examples, the corrections have a significant effect on the harmonic spectrum. The pseudo-Jahn-Teller $\tilde{B}$ and $\tilde{C}$ states, which are neglected here, also make a small contribution to the absorption cross section at the low energy part of the band. While accounting for these states requires a much more elaborate treatment~\cite{Ndengue2021}, it appears that the cubic/linear Herzberg-Teller simulation of the $\tilde{A}-\tilde{X}$ system alone accounts quantitatively for the gross features of the spectrum.

Because the $\tilde{A}$ state is still being treated harmonically, the auto-correlation function exhibits a regular series of revivals (Fig.~\ref{fig:NO2expt}B). The corresponding high-resolution spectrum (85~cm$^{-1}$ FWHM, Fig.~\ref{fig:NO2expt}C) contains extended vibrational progressions. As discussed further below, these high-resolution features are not expected to be quantitatively meaningful at this level of approximation, but have been included here for illustrative purposes. 

\section{Discussion}

The parameters of a vibronic spectrum are naturally partitioned into nearly independent contributions from the vertical electronic energy, the transition dipole, and their respective force-fields or derivatives. As the examples above show, when the potential energy surfaces of interest are approximately quadratic, the spectral contours can be accurately reproduced by analytical harmonic correlation functions augmented by low-order anharmonic and Herzberg-Teller corrections. In this limit, the remaining disagreement with measurements appears to be dominated by purely electronic errors in the vertical energy and oscillator strength. We stress that in dissociative or quasi-unbound spectra, what might normally be categorized separately as ``energy'' versus ``intensity'' effects are intrinsically convoluted. An accurate prediction of even basic descriptors, such as the peak spectral wavelength, require a balanced treatment of both. Indeed, this convolution poses a challenge to extensive quantum chemical benchmarking of vertical energies and oscillator strengths. A relatively inexpensive vibrational treatment as presented here may help bridge this gap, at least for those systems that possess simple enough potential energy surfaces in the vertical region.

The applications above have focused on systems with unstructured spectra, the envelopes of which are determined by the initial decay of the correlation function. This is not an inherent limitation of the time-dependent approach. In specific cases of conventional bound-to-bound transitions where the regions of the PES accessible to the propagating wavepacket are accurately described in the harmonic approximation, the long-time behavior of the correlation function is reliable, and in all cases the frequency-domain and time-domain results will of course be equivalent. When the vertical geometry is highly displaced from the upper-state minimum, however, the long-time behavior of the harmonic correlation function will not be accurate. The $\tilde{A}$ state of NO$_2$ examined above is a good example of this. Here, the vertical geometry is displaced from the upper state minimum by approximately $\Delta Q = 10$ in terms of the dimensionless bending normal coordinate. Although the harmonic correlation function exhibits partial revivals at the vibrational periods of the harmonic normal modes, these revivals are not meaningful given that between them the wavepacket traverses an extended region of the PES poorly described by the harmonic expansion. The same problem effectively occurs in a harmonic frequency-domain simulation. Accurate long-time simulations may be possible using time-dependent perturbation theory if the relevant region of the PES is well approximated with low-order anharmonic corrections. Bound vibrational modes, in particular, will probably require some sort of renormalized perturbative expansion of the correlation function~\cite{Iso2018,Note3}. \footnotetext[3]{In fact, the use of a vertical force field, rather than that of the upper-state minimum, is in some sense a crude form of renormalization.}

Finally, while we have focused on electronic absorption spectra, these methods are applicable to other techniques where FC or low-order Herzberg-Teller approximations are routinely employed, including photodetachment, photoionization, and photoelectron spectroscopy~\cite{Ruscic1986}, as well as non-radiative processes~\cite{Siebrand1967,Fischer1970,Peng2007}. It may also be possible to incorporate directly rotational effects such as origin shifts or axis-switching~\cite{Hougen1965}, which are important mostly for light molecules, especially small hydrides. Other terms in the rovibrational Hamiltonian, such as Coriolis coupling, should naturally be considered as higher-order perturbative anharmonic corrections to this framework are developed. 

\section{Conclusions}

We have presented a new derivation of the multidimensional harmonic oscillator auto-correlation function focused on the analytical and numerical issues associated with branch-cuts and stable finite-precision arithmetic. Together with the lowest-order anharmonic and Herzberg-Teller corrections, these results provide a simple means to simulating vibronic spectra that access unbound or negative curvature potential energy surfaces. The development of higher-order time-dependent perturbation theory corrections presents a promising avenue for future work.

\begin{acknowledgments}
This project has received funding from the European Research Council (ERC) under the European Union’s Horizon 2020 research and innovation programme (grant agreement  No.~848668). This work was also supported by the Israel Science Foundation (ISF), grant No.~194/20, Grant No.~2020724 from the United States-Israel Binational Science Foundation (BSF) and the US National Science Foundation (NSF award PHY-2110489).
\end{acknowledgments}

\section*{Conflict of Interest}
The authors declare no conflicts of interest.

\section*{Author's Contributions}
All authors contributed equally to this work.

\section*{Data Availability}
The data that support the findings of this study are available from the corresponding author upon reasonable request.

\appendix

\section{\label{app:disentangle}Derivation of Eqs.~\ref{eq:expAp} and \ref{eq:hp}}

We first consider the purely quadratic sub-algebra represented by the matrices $\mu$, which have the general form
\begin{align*}
    \mu &= \left( \begin{array}{cc} \mathbf{A} & 2\mathbf{B} \\ -2\mathbf{C} & -\mathbf{A}^T \end{array}\right).
\end{align*}
This is simply the center block of the full matrix representation, Eq.~\ref{eq:Mdef}.
For an exponential argument $-iHt$, this simplifies to
\begin{align*}
    \mu &= \frac{-i t}{2}\left( \begin{array}{c@{\quad}c}\mathbf{W} + \mathbf{K} & -\mathbf{W} + \mathbf{K} \\ \mathbf{W} - \mathbf{K} & -\mathbf{W} - \mathbf{K} \end{array}\right),
\end{align*}
where $\mathbf{W}$ and $\mathbf{K}$ are defined in Eq.~\ref{eq:Hdef}. This matrix has right eigenvectors
\begin{align*}
    \mathcal{L} &= \frac{1}{2}  \left(\begin{array}{cc}
\mathbf{\Lambda}_+ & -\mathbf{\Lambda}_- \\ -\mathbf{\Lambda}_- & \mathbf{\Lambda}_+ \end{array}\right)
\end{align*}
and left eigenvectors
\begin{align*}
    \mathcal{L}^{-1} &= \frac{1}{2} \left(\begin{array}{cc}
    \mathbf{\Lambda}_+^T & \mathbf{\Lambda}_-^T \\ \mathbf{\Lambda}_-^T & \mathbf{\Lambda}_+^T \end{array}\right),
\end{align*}
so that $\mu$ can be diagonalized as 
\begin{align*}
    \mu &= -it \times \mathcal{L}  \left(\begin{array}{cc}
\mathbf{\Sigma} \mathbf{\Omega} & 0 \\ 0 & -\mathbf{\Sigma}\mathbf{\Omega} \end{array}\right)  \mathcal{L}^{-1}.
\end{align*}
This permits its exponential to be simply expressed as 
\begin{align}
    e^\mu &= \mathcal{L}  \left(\begin{array}{cc}
e^- & 0 \\ 0 & e^+ \end{array}\right)  \mathcal{L}^{-1},\label{eq:expmu}
\end{align}
where $e^\pm \equiv \exp[\pm i \mathbf{\Sigma \Omega} t]$.

The full matrix representation of the general quadratic Hamiltonian, including gradient and constant terms, is
\begin{align*}
M &= \left( \begin{array}{ccc}
0 &  \left[ \begin{array}{cc} \mathbf{g}^T & \mathbf{f}^T\end{array}\right] & -2h \\
0 & \mu & \left[\begin{array}{c} -\mathbf{f} \\ \mathbf{g} \end{array}\right] \\
0 & 0 & 0
\end{array}\right).
\end{align*}
Repeated multiplication shows that higher powers of this matrix are
\begin{align*}
    M^k &= \left( \begin{array}{c@{\quad}c@{\quad}c}
0 & \left[ \begin{array}{cc} \mathbf{g}^T & \mathbf{f}^T\end{array}\right]\mu^{k-1} &\left[ \begin{array}{cc} \mathbf{g}^T & \mathbf{f}^T\end{array}\right]\mu^{k-2}\left[\begin{array}{c} -\mathbf{f} \\ \mathbf{g} \end{array}\right] \\
0 & \mu^k & \mu^{k-1}\left[\begin{array}{c} -\mathbf{f} \\ \mathbf{g} \end{array}\right] \\
0 & 0 & 0
\end{array}\right)
\end{align*}
for $k\geq 2$. By direct summation, the matrix exponential is therefore
\begin{align}
    e^M &= \left( \begin{array}{c@{\qquad}c@{\qquad}c}
1 &\left[ \begin{array}{cc} \mathbf{g}^T & \mathbf{f}^T\end{array}\right]\mu^{-1}(e^\mu - 1) &\left[ \begin{array}{cc} \mathbf{g}^T & \mathbf{f}^T\end{array}\right]\mu^{-2}(e^\mu - \mu - 1)\left[\begin{array}{c} -\mathbf{f} \\ \mathbf{g} \end{array}\right] - 2h \\
0 & e^\mu & \mu^{-1}(e^\mu - 1)\left[\begin{array}{c} -\mathbf{f} \\ \mathbf{g} \end{array}\right] \\
0 & 0 &1 
\end{array}\right).\label{eq:expM}
\end{align}

We now consider the disentangled operator, $e^{h'}e^{f'_i a_i^\dagger} e^{B'_{ij} a_i^\dagger a_j^\dagger} e^{A'_{ij}(a_i^\dagger a_j + \delta_{ij}/2)} e^{C'_{ij} a_i a_j} e^{g'_i a_i}$.
The matrix representation of each exponential factor can be evaluated using Eq.~\ref{eq:expM}. We assume a Hermitian Hamiltonian, for which $\mathbf{B}' = \mathbf{C}'$, $\mathbf{A}' = \mathbf{A}'^T$, and $\mathbf{g}' = \mathbf{f}'$. After matrix multiplication, the final product is
\begin{align}
    \left( \begin{array}{c@{\qquad}c@{\qquad}c}
1 & \left[ \begin{array}{c@{\quad}c} \mathbf{g}'^T -2\mathbf{g}'^T e^{-\mathbf{A}'} \mathbf{B}' & \mathbf{g}'^T e^{-\mathbf{A}'} \end{array}\right] & \mathbf{g}'^T e^{-\mathbf{A}'}\mathbf{g}' - 2h' \\
0 & \left[ \begin{array}{c@{\quad}c} e^{\mathbf{A}'} - 4\mathbf{B}'e^{-\mathbf{A}'}\mathbf{B}' & 2\mathbf{B}'e^{-\mathbf{A}'} \\ -2e^{-\mathbf{A}'}\mathbf{B}' & e^{-\mathbf{A}'} \end{array}\right] & \left[\begin{array}{c} -\mathbf{g}' + 2\mathbf{B}' e^{-\mathbf{A}'}\mathbf{g}' \\ e^{-\mathbf{A}'}\mathbf{g}' \end{array}\right] \\
0 & 0 &1 
\end{array}\right). \label{eq:disentangle_mat}
\end{align}
Equating Eq.~\ref{eq:expM} and Eq.~\ref{eq:disentangle_mat} allows us to solve for the disentangled parameters block-by-block. For example, the lower-right element of the central $2n \times 2n$ block yields
\begin{align*}
e^{-\mathbf{A}'} &= \frac{1}{4} \left( \mathbf{\Lambda}_+ e^+ \mathbf{\Lambda_+}^T - \mathbf{\Lambda}_- e^- \mathbf{\Lambda}_-^T \right),
\end{align*}
where $e^\mu$ has been expanded using Eq.~\ref{eq:expmu}. The matrix inverse of this expression gives Eq.~\ref{eq:expAp}.
Similarly, equating the upper-right elements and solving for $h'$ gives Eq.~\ref{eq:hp}.

\section{\label{app:finiteT}Derivation of the finite-temperature trace}

Beginning with the symmetric form of the thermal trace, Eq.~\ref{eq:splittrace}, the matrix representation of each factor is directly exponentiated and then multiplied. The final product is
\begin{align} 
\left( \begin{array}{c@{\quad}c@{\quad}c}
1 & \frac{-it}{\sqrt{2}}\left[ \begin{array}{cc}\mathbf{G}^T & \mathbf{G}^T \end{array}\right]\mathcal{L}\underline{\eta} \mathcal{L}^{-1}\underline{\xi}^{1/2} & \frac{-t^2}{2} \left[ \begin{array}{cc}\mathbf{G}^T & \mathbf{G}^T \end{array}\right]  \mathcal{L}\underline{\zeta}\mathcal{L}^{-1}\left[\begin{array}{c} -\mathbf{G} \\ \mathbf{G} \end{array}\right] + 2 i t V_0 \\
0 &\underline{\xi}^{1/2} \mathcal{L}\underline{e}\mathcal{L}^{-1}\underline{\xi}^{1/2} & \frac{-it}{\sqrt{2}}\underline{\xi}^{1/2}\mathcal{L}\underline{\eta} \mathcal{L}^{-1}\left[\begin{array}{c} -\mathbf{G} \\ \mathbf{G} \end{array}\right] \\
0 & 0 &1
\end{array}\right),\label{eq:traceproduct}
\end{align}
where $\xi = \exp[\mathbf{W}\tau]$ and
\begin{align*}
    \underline{e} &= \left[ \begin{array}{cc}e^- & 0 \\ 0 & e^+ \end{array}\right] ,\\
    \underline{\eta} &= \left[ \begin{array}{cc}\eta^- & 0 \\ 0 & \eta^+ \end{array}\right] ,\\
    \underline{\zeta} &= \left[ \begin{array}{cc}\zeta^- & 0 \\ 0 & \zeta^+ \end{array}\right], \\
    \underline{\xi}^{1/2} &= \left[ \begin{array}{cc}\xi^{-1/2} & 0 \\ 0 & \xi^{1/2} \end{array}\right].
\end{align*}
We now note a useful relation that allows us to evaluate $h'$, defined in Eq.~\ref{eq:hp_finiteT}, using only elements of the exponentiated representation, Eq.~\ref{eq:expM}, not the quadratic operator argument itself.
First, subtract identity from $e^\mu$ and invert, then sandwich this between the upper and right edges of the exponential representation. Finally, subtract the upper-right corner element and divide by 2. The result is
\begin{align*}
    \frac{1}{2}\left[ \begin{array}{cc} \mathbf{g}^T & \mathbf{g}^T\end{array}\right]\mu^{-1}\left[\begin{array}{c} -\mathbf{g} \\ \mathbf{g} \end{array}\right] + h &= - \mathbf{g}^T (\mathbf{A} + 2\mathbf{B})^{-1} \mathbf{g} + h\\
&= h'.
\end{align*}
Applying this procedure to Eq.~\ref{eq:traceproduct} yields
\begin{align*}
    h' &= -i t V_0 + \frac{t^2}{4} \left[ \begin{array}{cc}\mathbf{G}^T & \mathbf{G}^T \end{array}\right]\mathcal{L}\left\{\underline{\zeta}  -  \underline{\eta} \left(\underline{e} - \mathcal{L}^{-1}\underline{\xi}^{-1} \mathcal{L} \right)^{-1} \underline{\eta} \right\}\mathcal{L}^{-1}\left[\begin{array}{c} -\mathbf{G} \\ \mathbf{G} \end{array}\right].
\end{align*}

The regularization of the gradient terms starts with expanding this expression for $h'$.
The matrix inversion is handled block-wise assuming the bottom-right block and its respective Schur complement are invertible.
Then the matrix multiplication is carried out and terms are recollected as single-block expressions. After lengthy algebra, the result has the same form as Eq.~\ref{eq:hp},
\begin{align*}
h' = -itV_0 + \frac{t^2}{16} \mathbf{G}^T (\mathbf{\Lambda}_+ - \mathbf{\Lambda}_-) \mathbf{\Gamma} (\mathbf{\Lambda}_+ - \mathbf{\Lambda}_-)^T\mathbf{G},
\end{align*}
where 
\begin{subequations}
\begin{align}
\mathbf{\Gamma} &= (\zeta^+ - \zeta^-) - \theta_+^T\bar{D}^{-1} \theta_+  \label{eq:GammaT_zeroLimit}\\
&\qquad + \left( \theta_- - \bar{C}^T \bar{D}^{-1} \theta_+ \right)^T \xi^{-1/2} \left(\bar{A} + \xi^{-1/2} \bar{C}^T \bar{D}^{-1} \bar{C} \xi^{-1/2} \right)^{-1} \xi^{-1/2} \left( \theta_- - \bar{C}^T \bar{D}^{-1} \theta_+ \right)
\end{align}
\end{subequations}
and
\begin{align*}
\theta_\pm &=\mathbf{\Lambda}_\pm  \eta^+  + \mathbf{\Lambda}_\mp \eta^-  \\
\bar{A} &= \xi^{-1/2} ( \mathbf{\Lambda}_+ e^- \mathbf{\Lambda}_+^T - \mathbf{\Lambda}_- e^+ \mathbf{\Lambda}_-^T) \xi^{-1/2} - 4 \\
\bar{C} &= \mathbf{\Lambda}_+ e^+ \mathbf{\Lambda}_-^T - \mathbf{\Lambda}_- e^- \mathbf{\Lambda}_+^T\\
\bar{D} &= \mathbf{\Lambda}_+ e^+ \mathbf{\Lambda}_+^T - \mathbf{\Lambda}_- e^- \mathbf{\Lambda}_-^T - 4 \xi^{-1} 
\end{align*}

In the zero temperature limit, $\xi^{-1} \rightarrow 0$, only the first line of $\mathbf{\Gamma}$ remains (Eq.~\ref{eq:GammaT_zeroLimit}), and it reduces to the vacuum expectation value expression above. Its regularization is handled similarly,
\begin{align*}
(\zeta^+ - \zeta^-) - \theta_+^T\bar{D}^{-1} \theta_+  &= (1) + (2) + (3)\\
(1) &= \eta^+ \eta^- - \eta^+ \mathbf{\Lambda}_+^T (\mathbf{\Lambda}_+ e^+ \mathbf{\Lambda}_+^T - \mathbf{\Lambda}_- e^- \mathbf{\Lambda}_-^T - 4 \xi^{-1} )^{-1} \mathbf{\Lambda}_+ \eta^+\\
&=  \eta ^- e^+ \left[ 1  - \mathbf{\Lambda}_+^T (\mathbf{\Lambda}_+ e^+ \mathbf{\Lambda}_+^T - \mathbf{\Lambda}_- e^- \mathbf{\Lambda}_-^T - 4 \xi^{-1} )^{-1} \mathbf{\Lambda}_+ e^+ \right] \eta^- \\
&=  \eta ^- e^+ \left[ 1  -  (1 - e^- \mathbf{\Lambda}_+^{-1} ( \mathbf{\Lambda}_- e^- \mathbf{\Lambda}_-^T + 4 \xi^{-1}) \mathbf{\Lambda}'_+ )^{-1} \right] \eta^- \\
&=  \eta ^- e^+ \left[ ( - e^- \mathbf{\Lambda}_+^{-1} ( \mathbf{\Lambda}_- e^- \mathbf{\Lambda}_-^T + 4 \xi^{-1}) \mathbf{\Lambda}'_+ )  (1 - e^- \mathbf{\Lambda}_+^{-1} ( \mathbf{\Lambda}_- e^- \mathbf{\Lambda}_-^T + 4 \xi^{-1}) \mathbf{\Lambda}'_+ )^{-1} \right] \eta^- \\
&= -\eta ^- \mathbf{\Lambda}_+^{-1} ( \mathbf{\Lambda}_- e^- \mathbf{\Lambda}_-^T + 4 \xi^{-1}) \mathbf{\Lambda}'_+   (1 - e^- \mathbf{\Lambda}_+^{-1} ( \mathbf{\Lambda}_- e^- \mathbf{\Lambda}_-^T + 4 \xi^{-1}) \mathbf{\Lambda}'_+ )^{-1} \eta^- \\
(2) &= -2 \zeta^{-1} - \eta^- \mathbf{\Lambda}_-^T \bar{D}^{-1} \mathbf{\Lambda}_- \eta^-  \\
(3) &= -\eta^+ \mathbf{\Lambda}_+^T ( \mathbf{\Lambda}_+ e^+ \mathbf{\Lambda}_+^T - \mathbf{\Lambda}_- e^- \mathbf{\Lambda}_-^T - 4 \xi^{-1} )^{-1} \mathbf{\Lambda}_- \eta^- + \text{transpose} \\
&= -\eta^-  ( 1 - \mathbf{\Lambda}_+^{-1} (\mathbf{\Lambda}_- e^- \mathbf{\Lambda}_-^T + 4 \xi^{-1} )\mathbf{\Lambda}'_+ e^- )^{-1} \mathbf{\Lambda}_+^{-1} \mathbf{\Lambda}_- \eta^- + \text{transpose} 
\end{align*}
The remaining terms are
\begin{align*}
\theta_- - \bar{C}^T \bar{D}^{-1} \theta_+ &=  (1) + (2) + (3) + (4)\\
(1) &= \mathbf{\Lambda}_- \eta^+ - \mathbf{\Lambda}_- e^+ \mathbf{\Lambda}_+^T(\mathbf{\Lambda}_+ e^+ \mathbf{\Lambda}_+^T - \mathbf{\Lambda}_- e^- \mathbf{\Lambda}_-^T - 4 \xi^{-1} )^{-1} \mathbf{\Lambda}_+ \eta^+ \\
&= \mathbf{\Lambda}_-\left[ 1 - (1 - \mathbf{\Lambda}_+^{-1}(\mathbf{\Lambda}_- e^- \mathbf{\Lambda}_-^T + 4 \xi^{-1}) \mathbf{\Lambda}'_+ e^-)^{-1} \right] \eta^+ \\
&= \mathbf{\Lambda}_- \left[ (1 - \mathbf{\Lambda}_+^{-1}(\mathbf{\Lambda}_- e^- \mathbf{\Lambda}_-^T + 4 \xi^{-1}) \mathbf{\Lambda}'_+ e^-)^{-1} (-1)\mathbf{\Lambda}_+^{-1}(\mathbf{\Lambda}_- e^- \mathbf{\Lambda}_-^T + 4 \xi^{-1}) \mathbf{\Lambda}'_+ e^- \right] \eta^+ \\
&= -\mathbf{\Lambda}_- (1 - \mathbf{\Lambda}_+^{-1}(\mathbf{\Lambda}_- e^- \mathbf{\Lambda}_-^T + 4 \xi^{-1}) \mathbf{\Lambda}'_+ e^-)^{-1} \mathbf{\Lambda}_+^{-1}(\mathbf{\Lambda}_- e^- \mathbf{\Lambda}_-^T + 4 \xi^{-1}) \mathbf{\Lambda}'_+ \eta^-\\
(2) &= \mathbf{\Lambda}_+ \eta^- + \mathbf{\Lambda}_+ e^- \mathbf{\Lambda}_-^T\bar{D}^{-1} \mathbf{\Lambda}_- \eta^- \\
(3) &= -\mathbf{\Lambda}_- e^+ \mathbf{\Lambda}_+^T (\mathbf{\Lambda}_+ e^+ \mathbf{\Lambda}_+^T - \mathbf{\Lambda}_- e^- \mathbf{\Lambda}_-^T - 4 \xi^{-1} )^{-1} \mathbf{\Lambda}_- \eta^- \\
&= -\mathbf{\Lambda}_- ( 1 - \mathbf{\Lambda}_+^{-1}(\mathbf{\Lambda}_- e^- \mathbf{\Lambda}_-^T + 4 \xi^{-1})\mathbf{\Lambda}'_+ e^- )^{-1} \mathbf{\Lambda}_+^{-1} \mathbf{\Lambda}_- \eta^- \\
(4) &= \mathbf{\Lambda}_+ e^- \mathbf{\Lambda}_-^T ( \mathbf{\Lambda}_+ e^+ \mathbf{\Lambda}_+^T - \mathbf{\Lambda}_- e^- \mathbf{\Lambda}_-^T - 4 \xi^{-1} )^{-1} \mathbf{\Lambda}_+ \eta^+ \\
&= \mathbf{\Lambda}_+ e^- \mathbf{\Lambda}_-^T \mathbf{\Lambda}'_+ ( 1 - e^- \mathbf{\Lambda}_+^{-1}(\mathbf{\Lambda}_- e^- \mathbf{\Lambda}_-^T + 4 \xi^{-1} ) \mathbf{\Lambda}'_+ )^{-1} \eta^-
\end{align*}
and
\begin{align*}
\bar{A} + \xi^{-1/2}& \bar{C}^T \bar{D}^{-1} \bar{C} \xi^{-1/2}  = (1) + (2) + (3) \\
(1) &= \xi^{-1/2}\left[ - \mathbf{\Lambda}_- e^+ \mathbf{\Lambda}_-^T + \mathbf{\Lambda}_- e^+ \mathbf{\Lambda}_+^T (\mathbf{\Lambda}_+ e^+ \mathbf{\Lambda}_+^T - \mathbf{\Lambda}_- e^- \mathbf{\Lambda}_-^T - 4 \xi^{-1} )^{-1} \mathbf{\Lambda}_+ e^+ \mathbf{\Lambda}_-^T \right] \xi^{-1/2} \\
&= \xi^{-1/2} \mathbf{\Lambda}_- e^+ \left[ -1 + \mathbf{\Lambda}_+^T (\mathbf{\Lambda}_+ e^+ \mathbf{\Lambda}_+^T - \mathbf{\Lambda}_- e^- \mathbf{\Lambda}_-^T - 4 \xi^{-1} )^{-1} \mathbf{\Lambda}_+ e^+  \right] \mathbf{\Lambda}_-^T\xi^{-1/2} \\
&= \xi^{-1/2} \mathbf{\Lambda}_- e^+ \left[ -1 +  (1 - e^- \mathbf{\Lambda}_+^{-1} (\mathbf{\Lambda}_- e^- \mathbf{\Lambda}_-^T + 4 \xi^{-1})\mathbf{\Lambda}'_+ )^{-1}  \right] \mathbf{\Lambda}_-^T\xi^{-1/2} \\
&= \xi^{-1/2} \mathbf{\Lambda}_- e^+ \left[ e^- \mathbf{\Lambda}_+^{-1} (\mathbf{\Lambda}_- e^- \mathbf{\Lambda}_-^T + 4 \xi^{-1})\mathbf{\Lambda}'_+(1 - e^- \mathbf{\Lambda}_+^{-1} (\mathbf{\Lambda}_- e^- \mathbf{\Lambda}_-^T + 4 \xi^{-1})\mathbf{\Lambda}'_+ )^{-1}  \right] \mathbf{\Lambda}_-^T\xi^{-1/2} \\
&= \xi^{-1/2} \mathbf{\Lambda}_- \mathbf{\Lambda}_+^{-1} (\mathbf{\Lambda}_- e^- \mathbf{\Lambda}_-^T + 4 \xi^{-1})\mathbf{\Lambda}'_+(1 - e^- \mathbf{\Lambda}_+^{-1} (\mathbf{\Lambda}_- e^- \mathbf{\Lambda}_-^T + 4 \xi^{-1})\mathbf{\Lambda}'_+ )^{-1}  \mathbf{\Lambda}_-^T\xi^{-1/2} \\
(2) &= \xi^{-1/2}\left[ \mathbf{\Lambda}_+ e^- \mathbf{\Lambda}_+^T + \mathbf{\Lambda}_+ e^- \mathbf{\Lambda}_-^T \bar{D}^{-1} \mathbf{\Lambda}_- e^- \mathbf{\Lambda}_+^T\right] \xi^{-1/2} - 4 \\
(3) &= - \xi^{-1/2} \mathbf{\Lambda}_- e^+ \mathbf{\Lambda}_+^T (\mathbf{\Lambda}_+ e^+ \mathbf{\Lambda}_+^T - \mathbf{\Lambda}_- e^- \mathbf{\Lambda}_-^T - 4 \xi^{-1} )^{-1} \mathbf{\Lambda}_- e^- \mathbf{\Lambda}_+^T \xi^{-1/2} + \text{transpose}\\
&= - \xi^{-1/2} \mathbf{\Lambda}_-  (1  - \mathbf{\Lambda}_+^{-1} (\mathbf{\Lambda}_- e^- \mathbf{\Lambda}_-^T + 4 \xi^{-1}) \mathbf{\Lambda}'_+ e^- )^{-1} \mathbf{\Lambda}_+^{-1} \mathbf{\Lambda}_- e^- \mathbf{\Lambda}_+^T \xi^{-1/2} + \text{transpose}
\end{align*}

The remaining factor of the thermal correlation function contains the purely quadratic part of the thermal trace and the initial state partition function, $Z_0$. Together, these are
\begin{align*}
    (*) &\equiv (-1)^n \frac{1}{\det [e^\mu - \mathbf{1} ] Z_0^2}\\
    &= \frac{(-1)^n}{Z_0^2}\det \left[ \underline{\xi}^{1/2} \mathcal{L}\underline{e} \mathcal{L}^{-1}\underline{\xi}^{1/2} - \mathbf{1} \right]^{-1}\\
&= \frac{(-1)^n}{Z_0^2} \det \left[ \mathcal{L}\underline{e} \mathcal{L}^{-1} - \underline{\xi}^{-1} \right]^{-1}
\end{align*}

The harmonic partition function is
\begin{align*}
Z_0(\beta) &= \prod_i \frac{1}{2 \sinh \omega_i \beta/2 }\\
&=  \prod_i e^{-\omega_i \tau/2} e^{-i \omega_i t /2}\frac{1}{1 - e^{-\omega_i \beta}}\\
&= Z'_0(\beta)   \det [ e^{ -\mathbf{W} \tau/2} e^{-i \mathbf{W} t /2} ],
\end{align*}
where $Z_0'$ is the partition function with the leading exponential factor removed, i.e. with respect to the zero-point energy. The purely quadratic contribution is then
\begin{align*}
(*) &= \frac{(-1)^n}{Z'^2_0} \det [ \xi e^{i \mathbf{W} t } ] \det \left[ \mathcal{L}\underline{e} \mathcal{L}^{-1} - \underline{\xi}^{-1} \right]^{-1}\\
&= \frac{(-1)^n}{Z'^2_0} e^{it \text{Tr}\mathbf{W}} \det \left[ \left[ \begin{array}{cc}\xi^{-1/2} & 0 \\ 0 & 1 \end{array}\right] \mathcal{L}\left[ \begin{array}{cc}e^- & 0 \\ 0 & e^+ \end{array}\right]\mathcal{L}^{-1} \left[ \begin{array}{cc}\xi^{-1/2} & 0 \\ 0 & 1\end{array}\right] - \left[ \begin{array}{cc}1 & 0 \\ 0 & \xi^{-1} \end{array}\right] \right]^{-1}\\
&= \frac{(-1)^n}{Z'^2_0} e^{i t \text{Tr}\mathbf{W}} \det \left[\begin{array}{cc} \frac{1}{4}\xi^{-1/2}(\mathbf{\Lambda}_+ e^- \mathbf{\Lambda}_+^T - \mathbf{\Lambda}_- e^+ \mathbf{\Lambda}_-^T)\xi^{-1/2} - 1 & -\text{transpose}\\ \frac{1}{4}(\mathbf{\Lambda}_+ e^+ \mathbf{\Lambda}_-^T - \mathbf{\Lambda}_- e^- \mathbf{\Lambda}_+^T) \xi^{-1/2} & \frac{1}{4}(\mathbf{\Lambda}_+ e^+ \mathbf{\Lambda}_+^T - \mathbf{\Lambda}_- e^- \mathbf{\Lambda}_-^T) - \xi^{-1} \end{array}\right]^{-1}\\
&= \frac{1}{Z'^2_0} e^{i t \text{Tr}\mathbf{W}} \det \left[\begin{array}{cc} 1 - \frac{1}{4}\xi^{-1/2}(\mathbf{\Lambda}_+ e^- \mathbf{\Lambda}_+^T - \mathbf{\Lambda}_- e^+ \mathbf{\Lambda}_-^T)\xi^{-1/2} & +\text{transpose}\\ \frac{1}{4}(\mathbf{\Lambda}_+ e^+ \mathbf{\Lambda}_-^T - \mathbf{\Lambda}_- e^- \mathbf{\Lambda}_+^T) \xi^{-1/2} & \frac{1}{4}(\mathbf{\Lambda}_+ e^+ \mathbf{\Lambda}_+^T - \mathbf{\Lambda}_- e^- \mathbf{\Lambda}_-^T) - \xi^{-1} \end{array}\right]^{-1}
\end{align*}
Before continuing, we can already see that in the low $T$ limit, $\xi^{-1} \rightarrow 0$ and the determinant reduces to the vacuum expectation value result.

We now factor this matrix determinant into that of the lower-right block and its Schur complement,
\begin{align*}
\det \left[ \begin{array}{cc} A & B \\ C & D \end{array} \right] &= \det(D) \det(A - B D^{-1} C).
\end{align*}
The first factor is
\begin{align*}
\det \left[ \frac{1}{4}(\mathbf{\Lambda}_+ e^+ \mathbf{\Lambda}_+^T - \mathbf{\Lambda}_- e^- \mathbf{\Lambda}_-^T) - \xi^{-1} \right] &= \det \left[ \frac{1}{4}\mathbf{\Lambda}_+ e^+ \mathbf{\Lambda}_+^T(1 - \mathbf{\Lambda}'_+ e^- \mathbf{\Lambda}_+^{-1} (\mathbf{\Lambda}_- e^- \mathbf{\Lambda}_-^T + 4\xi^{-1})) \right]\\
&= \det(\mathbf{\Lambda}_+/2)^2 e^{i t \text{Tr}[\mathbf{\Sigma\Omega}]} \det \left[1 - \mathbf{\Lambda}'_+ e^- \mathbf{\Lambda}_+^{-1} (\mathbf{\Lambda}_- e^- \mathbf{\Lambda}_-^T + 4\xi^{-1}) \right]
\end{align*}
As with the zero-temperature case, the last factor is calculated by multiplying its eigenvalues, each of which does not cross the negative real axis. The principal square root can then be taken without crossing branch-cuts.

The second factor, the Schur complement, is
\begin{align*}
&\det\left[1 - \frac{1}{4} \xi^{-1/2} \left(\mathbf{\Lambda}_+ e^- \mathbf{\Lambda}_+^T - \mathbf{\Lambda}_- e^+ \mathbf{\Lambda}_-^T \right. \right. \\
&\qquad \left. \left. + (\mathbf{\Lambda}_- e^+ \mathbf{\Lambda}_+^T - \mathbf{\Lambda}_+ e^- \mathbf{\Lambda}_-^T)(\mathbf{\Lambda}_+ e^+ \mathbf{\Lambda}_+^T - \mathbf{\Lambda}_- e^- \mathbf{\Lambda}_-^T - 4 \xi^{-1})^{-1}(\mathbf{\Lambda}_+ e^+ \mathbf{\Lambda}_-^T - \mathbf{\Lambda}_- e^- \mathbf{\Lambda}_+^T) \right) \xi^{-1/2} \right].
\end{align*}
In the $e^+ \rightarrow \infty$ limit, all diverging terms cancel exactly. To remove all instances of $e^+$, we evaluate the total expression as 
\begin{align*}
\det[1 - \frac{1}{4} \xi^{-1/2}(\cdots) \xi^{-1/2}],
\end{align*}
where 
\begin{align*}
(\cdots) &= (1) + (2) + (3),\\
(1) &= \mathbf{\Lambda}_- e^+ \mathbf{\Lambda}_+^T (\mathbf{\Lambda}_+ e^+ \mathbf{\Lambda}_+^T - \mathbf{\Lambda}_- e^- \mathbf{\Lambda}_-^T - 4 \xi^{-1})^{-1} \mathbf{\Lambda}_+ e^+ \mathbf{\Lambda}_-^T - \mathbf{\Lambda}_- e^+ \mathbf{\Lambda}_-^T \\
&= \mathbf{\Lambda}_- \mathbf{\Lambda}_+^{-1} (\mathbf{\Lambda}_- e^- \mathbf{\Lambda}_-^T  + 4\xi^{-1})  (1- \mathbf{\Lambda}'_+ e^- \mathbf{\Lambda}_+^{-1}(\mathbf{\Lambda}_- e^- \mathbf{\Lambda}_-^T + 4 \xi^{-1}))^{-1}\mathbf{\Lambda}'_+\mathbf{\Lambda}_-^T\\
(2) &= \mathbf{\Lambda}_+ e^- \mathbf{\Lambda}_-^T  (\mathbf{\Lambda}_+ e^+ \mathbf{\Lambda}_+^T - \mathbf{\Lambda}_- e^- \mathbf{\Lambda}_-^T - 4 \xi^{-1})^{-1} \mathbf{\Lambda}_- e^- \mathbf{\Lambda}_+^T+ \mathbf{\Lambda}_+ e^- \mathbf{\Lambda}_+^T \\ 
&=\mathbf{\Lambda}_+ e^- \left[ \mathbf{\Lambda}_-^T  (1 - \mathbf{\Lambda}'_+ e^- \mathbf{\Lambda}_+^{-1} (\mathbf{\Lambda}_- e^- \mathbf{\Lambda}_-^T + 4 \xi^{-1}))^{-1} \mathbf{\Lambda}'_+ e^- \mathbf{\Lambda}_+^{-1} \mathbf{\Lambda}_- e^-  + 1\right] \mathbf{\Lambda}_+^T\\
(3) &= -\mathbf{\Lambda}_+ e^- \mathbf{\Lambda}_-^T (\mathbf{\Lambda}_+ e^+ \mathbf{\Lambda}_+^T - \mathbf{\Lambda}_- e^- \mathbf{\Lambda}_-^T - 4 \xi^{-1})^{-1} \mathbf{\Lambda}_+ e^+ \mathbf{\Lambda}_-^T + \text{transpose}\\
&= -\mathbf{\Lambda}_+ e^- \mathbf{\Lambda}_-^T (1 - \mathbf{\Lambda}'_+ e^- \mathbf{\Lambda}_+^{-1}  (\mathbf{\Lambda}_- e^- \mathbf{\Lambda}_-^T + 4 \xi^{-1}) )^{-1}\mathbf{\Lambda}'_+ \mathbf{\Lambda}_-^T + \text{transpose}
\end{align*}
We again assert that the eigenvalues of this matrix do not cross the negative real axis, which provides for a branch-cut free factorization of the principal square root.

\bibliography{references}

\end{document}